\documentstyle[12pt]{article}
\begin{document}
\newcommand{\beq}{\begin{equation}}
\newcommand{\eeq}{\end{equation}}
\newcommand{\dsp}{\displaystyle}
\newcommand{\ap}{\alpha}
\renewcommand{\theequation}{\arabic{equation}}
\vspace{1cm}

\hfill DSF-T-93/06
\hfill INFN -NA-IV-93/06

\begin{center}
\vskip 3cm
{ \LARGE \bf Correlation functions of quantum q-oscillators}
\vskip 1.8cm
{\Large V.I.Man'ko \footnote{on leave from Lebedev Physical Institute, Moscow},
G.Marmo, S.Solimeno, F.Zaccaria\\ Dipartimento di Scienze Fisiche\\
Univ. di Napoli Federico II \\ and I.N.F.N., Sez. di Napoli}\\
\vskip 0.5cm
\end{center}

\vskip 3cm
\begin{abstract}
Nonlinearity of electromagnetic field vibrations described by q-oscillators
is shown to produce essential dependence of second correlation functions
on intensity and deformation of Planck distribution. Experimental tests
of such nonlinearity are suggested.
\end{abstract}

\setcounter{footnote}{1}

\newpage

In the framework of investigations on quantum groups [1] the special
case of the Heisenberg-Weyl algebra was first considered by Biedenharn
 [2] and Macfarlane [3]. They introduced "q-oscillator operators" starting
from a deformed commutation relation containing a dimensionless parameter
q. The physical sense of the quantun q-oscillator up to now was not
clarified. In [4] a generalization to several modes has been analyzed.
 An attempt to connect quantum q-oscillator with relativistic
oscillator model has been discussed in [5]. The q-oscillators and quantum
group $SU_{q}(2)$ have also been considered in the frame of generalized
James-Cummings model [6]. Nevertheless these attempts are based on purely
mathematical properties of these operators.
In [7] aspects of the physical nature of q-oscillator have been considered
by the present authors. It was seen that it is subject to
 nonlinear vibrations with a special kind of
dependence of the frequency on the amplitude and the influence of such
nonlinearity on Bose distribution function was evaluated. Very recentlya
deformation of Bose distribution function has been studied also in [8].
The aim of this letter is to study the influence of the nonlinearity on
the correlation function of q-oscillator in different states, namely
coherent and squeezed ones and in a thermal bath.
We will show that such quantities acquire a dependence on the momenta of
the number operator. Assuming that the electromagnetic field be described
in terms of q-oscillators we will discuss the physical consequences of
this suggestion, namely how the presence of the deformation parameter
q changes the second order correlation formulae.
In [7] it was shown that the classical q-oscillator of amplitude $|\alpha|$
vibrates with a frequency
\beq
\omega_{q} = \omega \frac{\kappa \cosh \kappa |\alpha|^2}{\sinh\kappa}
\eeq
with $\omega\in {\bf R}$ and $\kappa = log q$. For $q\rightarrow 1$ linearity
of vibration is recovered.
q-oscillator operators can be defined anyhow also in terms of the usual
harmonic oscillator variables $(a, a^{\dag})$
\begin{eqnarray}
a_{q} & =& a \sqrt \frac{\sinh \hat n \kappa}{\hat n \sinh \kappa}\\
a_{q}^{\dag} & = & \sqrt{\frac{\sinh \hat n \kappa}{\hat n \sinh \kappa}}
a^{\dag}
\end{eqnarray}
in which $\hat n = a^{\dag}a$. The quantum system under consideration
then continues to be described by such last variables $(a,a^{\dag})$, but it
evolves under a Hamiltonian which in dimensionless units $(\omega=1)$ reads
\beq
 H_{q}(\hat n) =\frac{1}{2}\frac{1}{\sinh \kappa}(\sinh\kappa(\hat n +1)
+ \sinh\kappa\hat n).
\eeq
In terms of $(a_{q},a^{\dag}_{q})$ it reads
\beq
H_{q}(\hat n) = \frac{1}{2}(a_{q}a_{q}^{\dag} + a_{q}^{\dag}a_{q}).
\eeq
It is easy to prove that for an arbitrary function $V(\hat n)$
 which can be represented as a series of powers of $\hat n$ the following
 string of relations holds
\beq
V(\hat n) a = a V(\hat n -1), ~~~~V(\hat n)a^{\dag} = a^{\dag}V(\hat n + 1).
\eeq
The definition of second correlation function we will use is
\beq
\gamma(t) = Tr \rho a^{\dag}(t)a(0)
\eeq
with $\rho$ standing for the density matrix. Other definitions exist
for such a quantity, for instance with a normalization factor, but
we take the above which is simpler, as we wish to give an idea of the
corrections due to q-deformation, namely due to evolution under $H_{q}$.
We have therefore, using the string relations,
\beq
\gamma(t) = Tr \rho e^{itH_{q}(\hat n)}a^{\dag}(0)e^{-itH_{q}(\hat n)}a(0)
 = Tr \rho e^{it(H_{q}(\hat n) - H_{q}(\hat n -1)}\hat n.
\eeq
We will suppose nonlinearity appearing only at high intensities, so that
the limiting process
$q\rightarrow 1 (\kappa \rightarrow 0)$ is physically relevant. We have
\beq
H_{q}(\hat n) - H_{q}(\hat n - 1) = \frac{1}{2 \sinh\kappa}(\sinh
\kappa(\hat n+1) + \sinh \kappa (\hat n -1)) \simeq 1 +
\frac{\kappa^{2}\hat n^{2}}{2}.
\eeq
In what follows we will take all quantities in such an approximation so that
\beq
\gamma(t) \simeq e^{it} Tr \rho (1 + it \frac{\kappa^{2}\hat n^{2}}{2})\hat n.
\eeq
This formula is correct for finite times, not considering yet the asymptotic
behaviour of (8) for large times.
We consider first the case of a coherent state $|\alpha>,  \alpha \in {\bf C}$,
for which the density matrix is
\beq
\rho^{(coh)} = e^{|\alpha|^2}\sum_{m,n}\frac{\alpha^{n}\alpha^{*m}}{\sqrt{
n!m!}}|n><m|
\eeq
and therefore
\beq
\gamma^{(coh)}(t) \simeq |\alpha|^2 e^{it}(1+\frac{it\kappa^2}{2}(|\alpha|^4 +
3|\alpha|^2 + 1)) .
\eeq
We next consider the q-oscillator in the squeezed vacuum state
\beq
|0_{s}> = e^{\frac{r}{2}(a^2 - a^{\dag2})}|0>,
\eeq
where $r$ is the squeezing parameter defined by the equations
\begin{eqnarray}
<0_{s}|\frac{(a+a^{\dag})^2}{2}|0_{s}> & = &  \frac{1}{2}e^{-r}\\
<0_{s}|\frac{(a-a^{\dag})^2}{2}|0_{s}> & = & - \frac{1}{2}e^{r}.
\end{eqnarray}
The density matrix is now
\beq
\rho^{(sq)} = |0_{s}><0_{s}|
\eeq
and the second order correlation function in the above approximation
is
\beq
\gamma^{(sq)}(t) \simeq e^{it}(n_{s} + \frac{it\kappa^2}{2}n^{3}_{s}),
\eeq
where
\beq
n_{s} = <0_{s}|\hat n|0_{s}> = \sinh^{2} r
\eeq
and
\beq
n_{s}^3 = <0_{s}|{\hat n}^3|0_{s}> = \sinh^{2}2r \cosh 2r + \frac{3}{2}
\sinh^{2}2r \sinh^{2}r +\sinh^{6}r .
\eeq
The density matrix for the thermal case is
\beq
\rho^{(th)} =
e^{\frac{-\beta}{2}(a_{q}^{\dag}a_{q} + a_{q}a_{q}^{\dag})}\frac{1}
{Z_q}
\eeq
with
\beq
Z_{q} = \sum_{n=0}^{\infty} \exp\left[-{\beta\over 2}{\sinh\kappa(n+1)+
\sinh\kappa n\over \sinh\kappa}\right]
\eeq
and the parameter $\beta = \hbar \omega /kT$, with $\hbar$ Planck constant,
$\omega$ frequency, $k$ Boltzmann constant and $T$ absolute temperature.
In the considered approximation we write
\beq
Z_{q} \simeq Z +  a\kappa^2
\eeq
where $Z$ is the partition function for the harmonic oscillator
\beq
Z = \frac{1}{2\sinh\frac{\beta}{2}}
\eeq
and the coefficient $a$ can be calculated [7] and has the value
\beq
a = - \frac{\beta Z}{12}(2\bar {n^3} + 3\bar {n^2} + \bar n) .
\eeq
Then
\beq
\rho^{(th)} \simeq \frac{e^{-\beta(\hat n +  \frac{1}{2})}}{Z} ( 1 - \kappa^2
(\frac{a}{Z}  + \frac{\beta}{12}(2{\hat n}^3 + 3{\hat n}^2 + \hat n)))
\eeq
and, inserting this in (10), we obtain
\beq
\gamma^{(th)}(t) \simeq e^{it}\{\bar n + \kappa^2[\frac{it}{2}\bar{n^3} +
\frac{\beta}{6}\bar{n^3}\bar n + \frac{\beta}{4}\bar{n^2}\bar n + \frac{\beta}
{12}{\bar n}^2 - \frac{\beta}{6}\bar{n^4} - \frac{\beta}{4}\bar{n^3} -
\frac{\beta}{12}\bar{n^2}]\}.
\eeq
All the above equations contain some momenta of $\hat n$
in Planck distribution
\begin{eqnarray}
\bar n &=& \frac{1}{e^{\beta}-1},\\
\bar{n^2} &=& \frac{e^{\beta}+1}{(e^{\beta}-1)^2}.\\
\bar{n^3} &=& \frac{4e^{\beta} + e^{2\beta} + 1}{(e^{\beta}-1)^3} ,\\
\bar{n^4} &=& \frac{e^{3\beta} + 11e^{2\beta} + 11 e^{\beta} + 1
}{(e^{\beta}-1)^4}.
\end{eqnarray}
For $t=0$ the second correlation function gives the mean value of number
of photons $(\bar n)_{q}$ and so for the thermal state we have the deformed
Planck distribution formula [7]
\beq
(\bar n)_{q} \simeq \frac {1}{e^{\frac{\hbar \omega}{kT}}-1} - \kappa^2
\frac{\hbar \omega}{kT}\frac{e^{\frac{3\hbar\omega}{kT}} + 4 e^{\frac{2\hbar
\omega}{kT}} + e^{\frac{\hbar \omega}{kT}}}{(e^{\frac{\hbar \omega}{kT}} - 1)^4
}.
\eeq
Here the first term is the usual Planck distribution formula and its
correction is proportional to the square of the nonlinearity parameter
$\kappa$.\\

As it was seen the classical one-dimensional q-oscillator
 is nothing else than a nonlinear oscillator
with a very specific type of the nonlinearity. Namely, its frequency depends
on its energy as hyperbolic cosine of the energy. As a consequence spectra
should show a blue-shift effect when intensity of photon beams increases.
Should this phenomenon be observed it would be possibly explained also by other
types of nonlinearity.

The second-order
correlation functions studied by us depend on the higher momenta of the Bose
distribution for the photon number operator. For squeezed states,
 it is remarkable the larger
dependence of the correlation function on the squeezing parameter, comparing
with the standard oscillator.
As it was seen, the discussed q-nonlinearity produces a correction to
Planck distribution formula and also this may be subjected to an experimental
test.
All these properties mark the main difference with the standard, linear
 electromagnetism.

Therefore, in spite of the mathematical beauty of q-deformation procedure,
 the most important issue is to envisage
 how to test experimentally the possibility of describing
electrodynamic phenomena by means of these q-oscillators.
At a first thought , non-linear optics
experiments appear as potential candidates for obtaining upper bounds for the
value of $\kappa$.
Another class of experiments could be based on the dependence of
time-correlation function  on the intensity. The interpretation
of these experiments does not depend critically on the development of a
q-field QED. In fact, these measurements can be carried out in vacuum. For
example, a continuous laser beam could be split in two beams of
different intensities  and sent to two spectra analyzers. Measuring the
widths of the two spectra it would be possible to put an upper
limit to the deformation parameter
$\kappa$. These measurements could last for years thus guaranteeing a
signal-to-noise
ratio adequate for appreciating very small values of the parameter $\kappa$.

\vskip 2cm
Acknowledgements.
One of us (V.I.M.) thanks INFN and University of Napoli "Federico II"
for the hospitality.

\newpage
\begin{center}
{\bf References}
\end{center}

\bigskip

\noindent
1. V.G. Drinfeld, Quantum Groups, {\sl Proc. Int. Conf.\ of Math.},
MSRI Berkeley CA (1986), p. 798.

\noindent
2. L. C. Biedenharn, {\sl J. Phys.} {\bf A22}, L873 (1989).

\noindent
3. A. J. Macfarlane, {\sl J. Phys.} {\bf A22}, 4581 (1989).

\noindent
4. D.B.Fairlie and C.K.Zachos, {\sl Phys.Lett.} {\bf B256}, 43 (1991).

\noindent
5. R. M. Mir-Kasimov, Proc. of the 18th Group Theoretical Methods in Physics
Colloquium, Moscow, June (1990) in Lecture Notes in Physics, v. 382,
p. 215, Springer (1991).

\noindent
6. M. Chaichian, D. Ellinas, P. Kulish, {\sl Phys. Rev. Lett.} {\bf 65}
980 (1990).

\noindent
7. V.I.Man'ko, G.Marmo, S.Solimeno, F.Zaccaria, Physical nonlinear aspects
of classical and quantum q-oscillators, Napoli preprint DSF-T-92/25 INFN
-NA-IV-92/25, to appear in Int.Jour.Mod.Phys.

\noindent
8. Gang Su and Mo-lin Ge, {\sl Phys.Lett.} {\bf A173}, 17 (1993).
\noindent

\end{document}